# Controllable phase transitions between multiple charge density waves in monolayer 1T-VSe$_2$ via doping and strain engineering

*Zishen Wang,$^{1,3}$ Jun Zhou,$^{1,*}$ Kian Ping Loh,$^{2,3,*}$ Yuan Ping Feng$^{1,3,*}$*

[1] Department of Physics, National University of Singapore, 117542, Singapore

[2] Department of Chemistry, National University of Singapore, 117543, Singapore

[3] Centre for Advanced 2D Materials, National University of Singapore, 117546, Singapore

\* Corresponding authors:

Jun Zhou (phyzjun@nus.edu.sg)

Kian Ping Loh (chmlohkp@nus.edu.sg)

Yuan Ping Feng (phyfyp@nus.edu.sg)

**ABSTRACT:** Two-dimensional (2D) materials are known to possess emergent properties that are not found in their bulk counterparts. Recent experiments have shown a $\sqrt{7} \times \sqrt{3}$ charge density wave (CDW) in monolayer 1T-VSe$_2$, in contrast to the $4 \times 4 \times 3$ phase in bulk. Here, via first-principles calculations, we show that multiple CDW phases compete in monolayer VSe$_2$, the ground state of which can be tuned by charge doping and in-plane biaxial strain. With doping, the $\sqrt{7} \times \sqrt{3}$ CDW of the pristine VSe$_2$ transfers to a $3 \times \sqrt{3}$ and $4 \times 4$ phase, the latter of which is a projection of the bulk counterpart, at critical doping concentrations of around 0.2 holes per formula unit and 0.25 electrons per formula unit, respectively. The $4 \times 4$ CDW phase can also be stabilized under compressive strain. Although electron-phonon coupling is prevailing in the CDW formation, we show that Fermi surface nesting is a good starting point to explain most of these transitions in monolayer 1T-VSe$_2$. These results make VSe$_2$ an appealing material for electronic devices based on controllable CDW phase transitions.

Since the discovery of graphene [1], the family of 2D materials has grown at an unprecedented speed [2,3]. Among them, monolayer transition metal dichalcogenides (TMDs) with chemical formula MX$_2$ (M = transition metal and X = chalcogen) have received tremendous attention due to their rich intriguing physical properties, such as ferromagnetism, superconductivity, and valleytronics [4–8] to name a few. In particular, charge density wave (CDW), as widely observed in metallic 2D TMDs, attracts broad research interest for its complex entanglement and competition with other quantum ordering phenomena [9,10]. The concept of CDW was first proposed by Peierls to describe the instability of one-dimensional metal at a low temperature [10]. This generates the prototype of the Fermi surface nesting picture, which has been complemented by electron-phonon coupling recently [11–13]. In both mechanisms, below a critical temperature ($T_C$), the electronic energy gain from a gap opening at some ***k*** points on the Fermi surface outweighs the elastic energy loss from the lattice distortions [12], and CDW occurs as a spontaneous symmetry-lowering process with charge redistribution and lattice distortions [11,14]. The delicate balance between the two energy terms makes the CDW order very sensitive to changes in its electron and phonon properties, providing great flexibility for manipulating CDW properties by external stimuli. Several experimental results have been reported along this line, for example: (1) Applying an in-plane electric field changes the commensurate charge density wave (CCDW) state of 1T-TaS$_2$ to a nearly commensurate charge density wave (NCCDW) state [15]; (2) Cu-intercalation can induce CDW incommensuration in 1T-TiSe$_2$, which facilitates the formation of superconductivity [16]; (3) strain has been shown to introduce changes in the CDW vector and geometry in 2H-NbSe$_2$ [17].

Recently, monolayer 1T-VSe$_2$ (in short VSe$_2$) has been successfully synthesized [18,19] and becomes a hot topic partially due to its observed room-temperature 2D ferromagnetism [20]. However, controversial experimental results on the existence of such 2D ferromagnetism in VSe$_2$ have been reported and a unified picture is lacking. Interestingly, CDW distortions have been proposed to suppress the ferromagnetism in pristine monolayer VSe$_2$ [18,19,21,22]. Nevertheless, the CDW properties of this interesting 2D material are largely unexplored and the underlying physics remains elusive. For example, a $4 \times 4 \times 3$ CDW structure has been reported in bulk VSe$_2$ with a transition temperature of around 110 K [23], while a different $\sqrt{7} \times \sqrt{3}$ CDW has been identified in VSe$_2$ monolayer by serval experiments [18,21,22,24]. Compared to the bulk VSe$_2$ CDWs, the broken C$_3$ symmetry in monolayer VSe$_2$ CDWs is unexpected considering the quasi-2D nature of TMDs. Moreover, the $T_C$ of the CDW in monolayer VSe$_2$ spans over a wide range from 121 K [20] to 350 K [24]. This suggests that the CDW in monolayer VSe$_2$ might be sensitive to ambient factors.

In this work, first-principles calculations were performed to study the multiple CDW phases of monolayer VSe$_2$ and possible manipulation of ground state CDW by charge doping and strain effects. It is found that the $\sqrt{7} \times \sqrt{3}$ CDW structure has the lowest energy in the pristine case, in line with the experimental observations, while a compressive strain or a sufficient electron doping leads to a preference of the $4 \times 4$ CDW. Besides, a new $3 \times \sqrt{3}$ phase, which has not been observed in the experiments for monolayer VSe$_2$, dominates under hole doping. Our results shed light on the multiple CDW phases of monolayer VSe$_2$ and their mutual transitions by charge doping and biaxial strain.

**Methods**

The calculations were performed using the Vienna *ab initio* simulation package (VASP) [25,26] with the projector augmented wave (PAW [27]) method. The Perdew-Burke-Ernzerhof (PBE) form of the generalized gradient approximation (GGA [28]) was chosen to describe the electron exchange and correlation effects. A cutoff energy of 550 eV and a *k*-mesh of $28 \times 28 \times 1$ over the Brillouin zone were employed based on the results of convergence tests. A vacuum gap of 15 Å was chosen to minimize the interaction between adjacent layers. The in-plane lattice parameter was fixed to the experimentally reported value of 3.35 Å [29]. All the atoms in the VSe$_2$ structures were fully optimized until ionic forces and energy differences were less than 0.001 eV/Å and $10^{-5}$ eV, respectively. Phonon calculations were performed using the density functional perturbation theory (DFPT) implemented in the Quantum Espresso (QE) [30,31] package. The energy cutoff was set to 80 Ry, and the phonon dispersion curves were calculated with a $16 \times 16 \times 1$ *k*-mesh. An $8 \times 8 \times 1$ *q*-point grid was applied to obtain the dynamic matrices. The Fermi-Dirac smearing method was employed to simulate the temperature effects. The charge doping was simulated by adding (electron doping) or removing (hole doping) electrons in the system with a compensating uniform positive (negative) background. The crystal structures were re-optimized without changing the lattice constant for each doping concentration. In the investigation of in-plane biaxial strain, the lattice constant $a_0$ was changed according to the strain $\varepsilon = (a - a_0)/a_0$ with atomic positions relaxed. Considering that CDW distortions suppress the ferromagnetic state as mentioned above, all CDW properties were simulated by using nonmagnetic calculations [32]. BandUP code was used to unfold the CDW band structures [33,34].

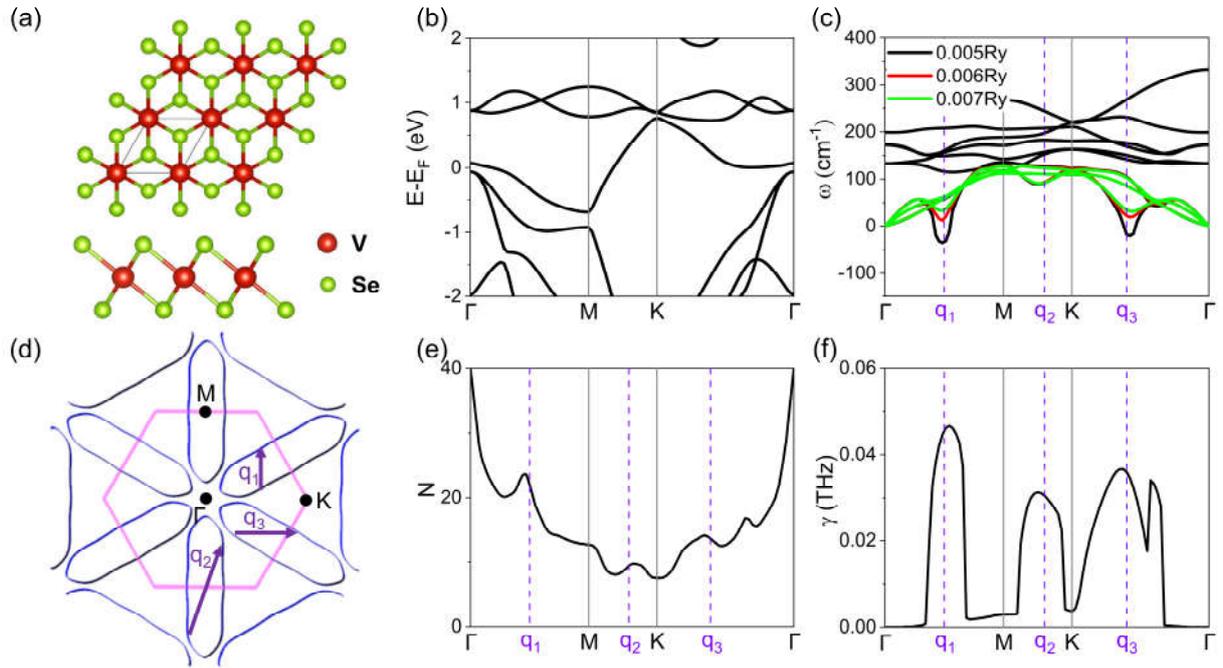

**Figure 1.** (a) Structure of the normal VSe$_2$ monolayer in top and side view. (b) Band structure for monolayer VSe$_2$. (c) Phonon dispersion of monolayer VSe$_2$ with different smearing factors σ. For clarity, we only show acoustic phonon modes for σ larger than 0.005 Ry. (d) Fermi surface in the Brillouin zone. Nesting vectors are indicated by purple arrows. (e) Nesting function N(**q**) and (f) phonon linewidth $\gamma$(**q**) of the soften phonon mode for monolayer VSe$_2$.

The structure of pristine monolayer VSe$_2$ is shown in Fig. 1(a), which has a P-$\bar{3}$m1 space group and consists of three sublayers with V atoms sandwiched between Se layers. In the following, we refer the pristine VSe$_2$ without CDW as *normal* VSe$_2$. As the band structure is shown in Fig. 1(b), monolayer VSe$_2$ is metallic, which is a prerequisite of the formation of CDW, and there is an electron pocket at the *M* point, in good agreement with previous experimental and theoretical results [18,22]. The phonon dispersions of normal VSe$_2$ with different smearing factors σ are shown in Fig. 1(c). The smearing factor under the Fermi-Dirac distribution in the metallic system can indicate the electronic temperature [35]. It is noticeable that the lowest acoustic phonon branch is softened at two **q** points, $\mathbf{q_1} = \frac{1}{2}\Gamma M$ and $\mathbf{q_3} = \frac{3}{5}\Gamma K$, with the decrease of temperature and becomes imaginary at the temperature corresponding to σ = 0.005 Ry, demonstrating sharp soft modes or Kohn anomalies [36]. These imaginary frequencies indicate that the structure of the normal phase is unstable at low temperature and may promote CDW phase transitions.

The **q** vectors in the first Brillouin zone (BZ) of these imaginary phonon frequencies can be converted to the real space lattice vectors of the potential CDW structures [18,32]. For example, the $\mathbf{q_1}$ vector corresponds to a 4 × 4 CDW as shown in Figs. 2(a)-2(b) and Fig. s2(a). And the combination of $\mathbf{q_3}$ and another vector $\mathbf{q_2} = \frac{3}{5}MK$, where the phonon dispersion shows a dip with the Kohn anomaly [see Fig. 1(c)], suggests a $\sqrt{7} \times \sqrt{3}$ CDW as shown in Figs. 2(c)-2(d) and Fig. s2(b). The simulated STM images of these two CDW phases [see Figs. s2(a)-s2(b)] match the experimental observations [18,37], which demonstrates the reliability of our CDW structures. And the partial CDW gap along the *MK* path as shown in the unfolded band structure of the $\sqrt{7} \times \sqrt{3}$ phase into the primitive BZ [Fig. s1] is also in good agreement with the experimental gap position [18,38]. The coexistence of these imaginary frequencies suggests that multiple structural instabilities compete in normal $VSe_2$ [18]. Energetically, the $\sqrt{7} \times \sqrt{3}$ CDW is 4 meV per formula unit (meV/f.u.) lower than the 4 × 4 phase, which explains the more frequently observed $\sqrt{7} \times \sqrt{3}$ CDW by experimental reports [18,21,22,24].

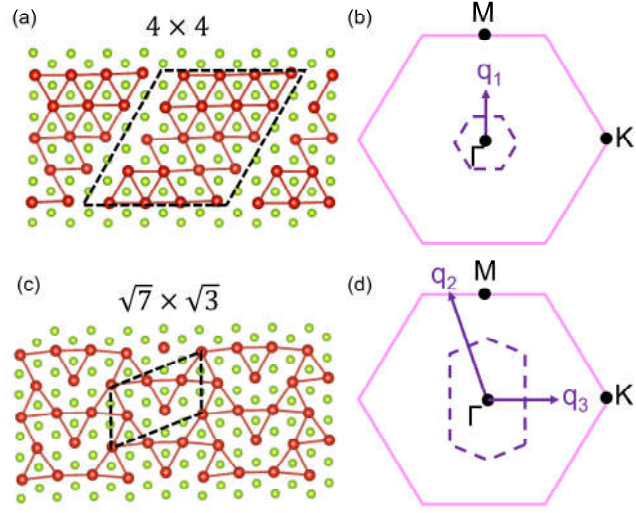

**Figure 2.** Schematic diagrams of (a) 4 × 4 CDW and (c) √7 × √3 CDW phases. The black dash parallelograms denote the periodic lattice distortions of CDW phases. (b) The first BZ of 4 × 4 CDW phase (purple dashed lines) with primitive reciprocal lattice vector $\mathbf{q_1}$ (purple arrow), and (d) the first BZ of √7 × √3 CDW phase (purple dashed lines) with primitive reciprocal lattice vectors $\mathbf{q_2}$ and $\mathbf{q_3}$. The first BZ of the normal VSe$_2$ is indicated in pink solid lines.

To understand the underlying mechanisms, the Fermi surface of monolayer normal VSe$_2$ is calculated. As shown in Fig. 1(d), there are six ellipse-shaped electron pockets, each of which is centered at one of the six *M* points of the first BZ. Such electron pockets provide an excellent condition for Fermi surface nesting [18,24,32]. Interestingly, the three CDW vectors $\mathbf{q_1}$, $\mathbf{q_2}$ and $\mathbf{q_3}$ are very close to perfect nesting condition, which is expected to give rise to 4 × 4 and √7 × √3 CDW structures, in line with the discussions above. For a quantitative understanding, the nesting function N($\mathbf{q}$), and phonon linewidth $\gamma(\mathbf{q})$ are calculated along the high symmetry path of normal VSe$_2$:

$$N(\mathbf{q}) = \int_{BZ} \frac{d\mathbf{k}}{\Omega_{BZ}} \delta(\varepsilon_{\mathbf{k}} - \varepsilon_F)\delta(\varepsilon_{\mathbf{k+q}} - \varepsilon_F) \qquad (1)$$

$$\gamma(\mathbf{q}) = 2\pi\omega(\mathbf{q}) \int_{BZ} \frac{d\mathbf{k}}{\Omega_{BZ}} |g(\mathbf{k},\mathbf{q})|^2 \delta(\varepsilon_{\mathbf{k}} - \varepsilon_F)\delta(\varepsilon_{\mathbf{k}+\mathbf{q}} - \varepsilon_F) \qquad (2)$$

where $\Omega_{BZ}$ is the area of BZ, $g(\mathbf{k},\mathbf{q})$ is the electron-phonon coupling matrix, $\varepsilon_{\mathbf{k}}$ and $\varepsilon_F$ are the band energy and Fermi energy, respectively. The nesting function is a direct measurement of Fermi surface nesting, which shows peaks at the nesting vectors [18], while the phonon linewidth takes into account both effects of electron-phonon coupling and Fermi surface nesting [13]. As shown in Figs. 1(e) and 1(f), the nesting function and phonon linewidth both produce peaks at around the three commensurate CDW vectors of the normal VSe$_2$ [18,39]. This supports the claim that Fermi surface nesting is sufficient to understand the CDW in monolayer VSe$_2$ [18,40]. However, it is noted that the CDWs can gain energy through lock-in transitions from their ordering vectors to the neighboring commensurate CDW vectors [41], which are expected in the realistic material [18,42]. Thus, in the following, we apply the Fermi surface nesting scenario as a simple starting point to understand the multiple CDW structures in monolayer VSe$_2$ [40].

Given the possibility of multiple competing CDW states in monolayer VSe$_2$, in the following, we use charge doping and biaxial strain to tune the energetic hierarchy of these CDWs, which is also helpful to shed light on the inconsistencies in experimental observations. Fig. 3 shows the results of the VSe$_2$ monolayer by charge doping. As shown in Fig. 3(a), electron doping does not change the overall profile of the phonon dispersions. However, there are slight shifts of the three acoustic phonon dips with doping concentration, that is, from near $\mathbf{q_1}, \mathbf{q_2}$ and $\mathbf{q_3}$ towards *M*, *K* and *K*, respectively. These shifts can be explained by the increase of Fermi energy under electron doping and the corresponding change of the Fermi surface [see Fig. 3(e)], which is similar to a typical $2k_F$ effect [40], implying the important role of the Fermi surface nesting mechanism played in monolayer VSe$_2$. By contrast, hole doping changes the profile of the acoustic phonon modes significantly [see Fig. 3(b)]. All the three dips of acoustic phonon in the normal phase are suppressed and new imaginary frequencies spike at $\mathbf{q_4} = M$ and $\mathbf{q_5} = \frac{1}{2}\Gamma K$ points [see the corresponding CDW vectors in Fig. s8(a)], the combination of which corresponds to a new $3 \times \sqrt{3}$ CDW phase that has not been reported in VSe$_2$ [see the structure and simulated STM in Figs. 3(c) and s2(c), respectively]. The CDW structures deduced from either $\mathbf{q_4}$ or $\mathbf{q_5}$ vector [see Figs. s9(a) and s9(b)] have higher energy than the $3 \times \sqrt{3}$ CDW, and are not considered in the following discussions. It is noted that the imaginary frequencies around the $M$ point spread over a large range along the high symmetry path, in contrast to a sharp dip in the Fermi surface nesting picture, and there is no corresponding nesting vector $\Gamma M$ on the Fermi surface [Figs. 3f and s6(a)]. However, a large and broad peak is showed in its phonon linewidth at the $M$ point [Fig. s6(b)], suggesting electron-phonon coupling as the driving force of the instability at the $M$ point of VSe$_2$ under a high hole doping concentration [11,40,43].

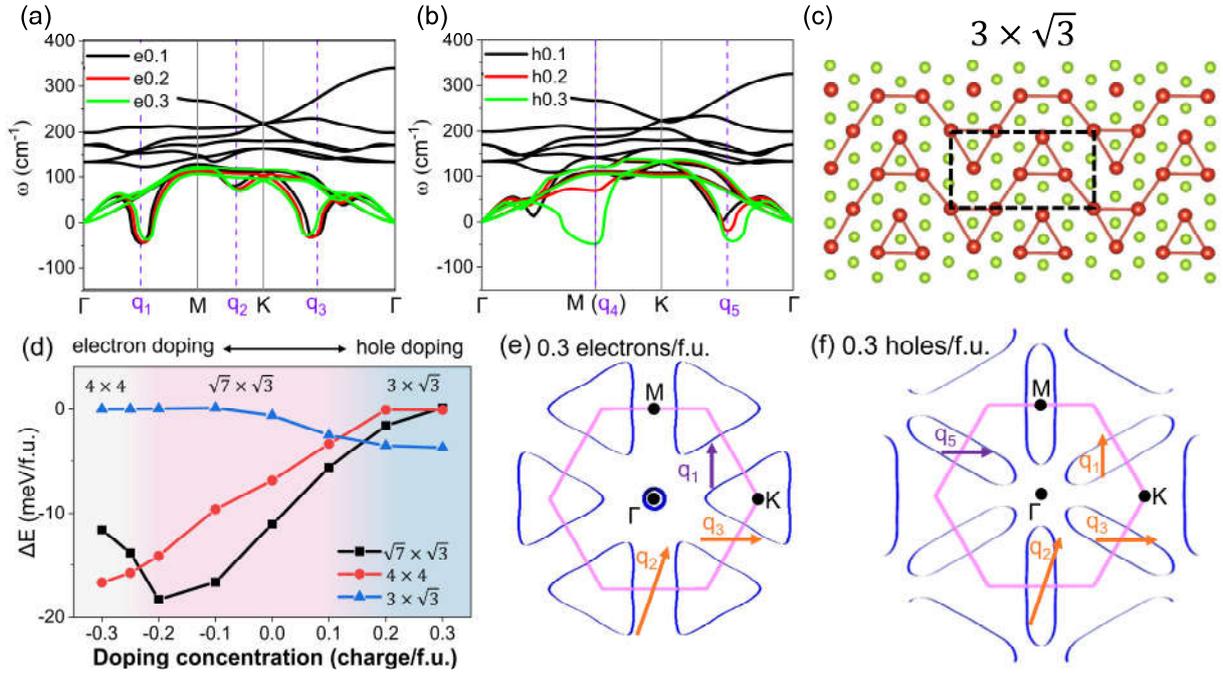

**Figure 3.** Evolution of phonon dispersion curves of monolayer VSe$_2$ with (a) electron doping and (b) hole doping. For clarity, we only show acoustic modes when doping concentration above 0.1 charges/f.u.. (c) Top view of $3 \times \sqrt{3}$ CDW structure. (d) The formation energies for the $\sqrt{7} \times \sqrt{3}$, $4 \times 4$, and $3 \times \sqrt{3}$ CDW structures as a function of doping concentration. The shade regions with different colors stand for different CDW ground state: light gray for $4 \times 4$, light pink for $\sqrt{7} \times \sqrt{3}$, and light blue for $3 \times \sqrt{3}$ CDW phase. Additional calculations have been performed for -0.25 charges/f.u. doping to confirm the phase transition around this concentration. (e) and (f) are Fermi surfaces of monolayer VSe$_2$ with 0.3 electrons/f.u. and 0.3 holes/f.u. doping concentrations, respectively. First BZs are indicated by pink solid lines. The nesting vectors under good (poor) nesting conditions are denoted by purple (orange) arrows.

To study the possible CDW phase transitions by charge doping, the formation energies of all the three CDW structures are calculated for different doping concentrations. The formation energy is defined as $\Delta E = E_{CDW}/x - E_0$, where $E_{CDW}$, $E_0$ and $x$ are the total energy of VSe$_2$ with CDW, the $1 \times 1$ normal phase, and the number of formula units in the CDW structure, respectively. Fig. 3(d) shows the dependence of the formation energies of VSe$_2$ for the $\sqrt{7} \times \sqrt{3}$, $4 \times 4$, and $3 \times \sqrt{3}$ CDW structures on the doping concentrations. Overall, their formation energies are negative, indicating their favorable formation in monolayer VSe$_2$, except for the $\sqrt{7} \times \sqrt{3}$ and $4 \times 4$ phases in the high hole doping range ($\geq 0.3$ charges/f.u.) as well as for the $3 \times \sqrt{3}$ CDW in the whole electron doping range studied in this work. Interestingly, electron doping stabilizes while hole doping suppresses the $4 \times 4$ CDW, and similar behavior can be observed for the $\sqrt{7} \times \sqrt{3}$ CDW structure except an upturn from 0.2 to 0.3 electrons/f.u. doping, while an opposite trend is demonstrated for the $3 \times \sqrt{3}$ CDW. These results are in line with the evolution of the phonon dispersions with charge doping. Overall, the $\sqrt{7} \times \sqrt{3}$ CDW structure dominates over a doping range from -0.2 to 0.1 charges/f.u at low temperature, in line with most of the experimental observations [18,21,22,24]. Remarkably, two phase transitions from $\sqrt{7} \times \sqrt{3}$ to $3 \times \sqrt{3}$ and to $4 \times 4$ occur around the hole doping concentration of 0.2 charges/f.u. and electron doping of 0.25 charges/f.u., respectively.

To further understand the phase transitions between different CDW structures, we calculated the band structures and Fermi surfaces for normal VSe$_2$ with a doping concentration of 0.3 electrons/f.u. and 0.3 holes/f.u.. As shown in Figs. 3(e) and s3(a), the 0.3 electrons/f.u. doping changes the Fermi surface of VSe$_2$ from the previous cigar-like electron pockets to nearly triangular hole pockets centered at the $K$ points of the first BZ, due to the raise of the Fermi level.

Besides, an extra band compared with the band structure of pristine monolayer VSe$_2$ appear around the Fermi level [44] and forms a small electron pocket at $\Gamma$ point as shown in the Fermi surface. It is noted that the nesting conditions for the $\mathbf{q_2}$ and $\mathbf{q_3}$ vectors, which correspond to the $\sqrt{7} \times \sqrt{3}$ CDW, are not satisfied in this doping concentration as shown by the highlighted orange arrows in Fig. 3(e) and the nesting function in Fig. s5(a). This indicates that the system will spend some energy to lock back to the lattice for $\sqrt{7} \times \sqrt{3}$ CDW, increasing the total energy of this state [40] [see Fig. 3(d)]. However, the $\mathbf{q_1}$ vector, which corresponds to the $4 \times 4$ CDW, still meets the nesting condition. This explains the lower formation energy of the $4 \times 4$ CDW than that of the $\sqrt{7} \times \sqrt{3}$ CDW at 0.3 electrons/f.u. doping.

For the case of 0.3 holes/f.u. doping, the decrease of the Fermi energy shrinks the 2D ellipse-shaped electron pockets, leading to the failure of the nesting condition for $\mathbf{q_1}$, $\mathbf{q_2}$ and $\mathbf{q_3}$ as the highlighted orange arrows are shown in Fig. 3(f). However, the shrinkage of the electron pockets creates an opportunity for the new $\mathbf{q_5}$ to meet the nesting condition under this doping concentration [see the purple arrow in Fig. 3(f)], which in accord with one of the $3 \times \sqrt{3}$ CDW nesting vectors. Combined with $\mathbf{q_4}$ at the $M$ point, the $3 \times \sqrt{3}$ CDW state becomes the ground state at this doping concentration.

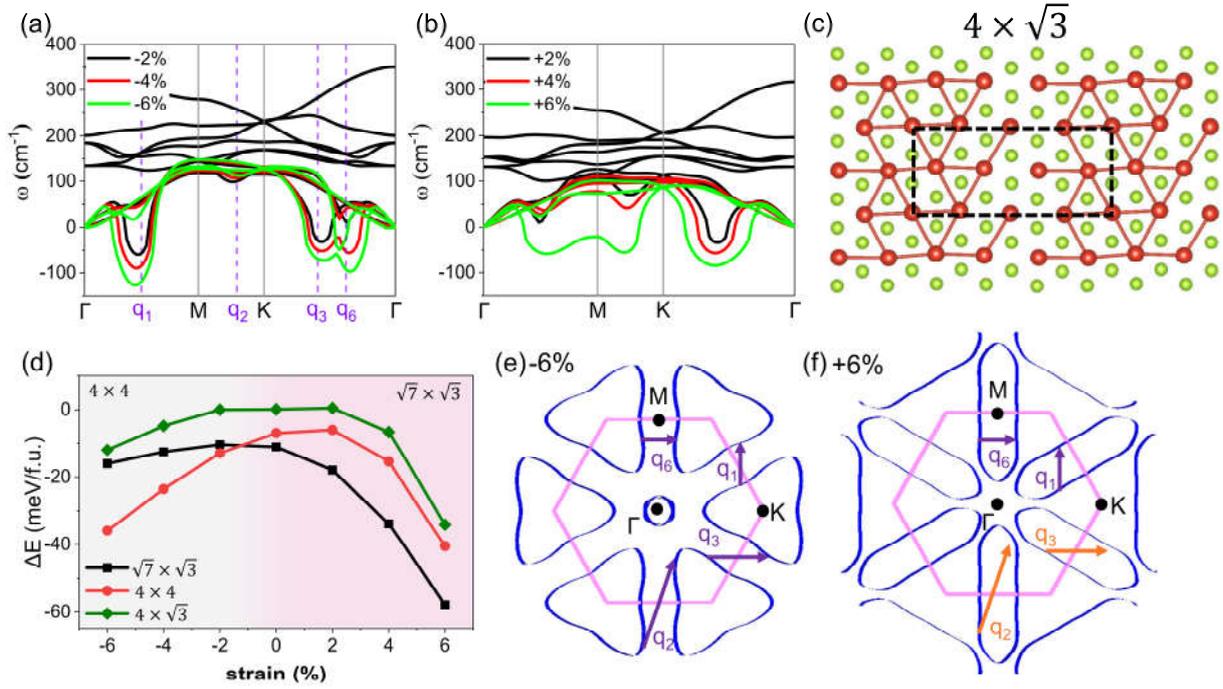

**Figure 4.** Evolution of phonon dispersion curves of monolayer normal VSe$_2$ under in-plane biaxial (a) compressive and (b) tensile strain. Only acoustic modes are shown for cases with strain (compressive or tensile) larger than 2%. (c) Top view of $4 \times \sqrt{3}$ CDW structure. (d) The formation energies for the $\sqrt{7} \times \sqrt{3}$, $4 \times 4$, and $4 \times \sqrt{3}$ CDW as a function of biaxial strain. The shaded region with light gray stands for the $4 \times 4$ CDW structure, while light pink for the $\sqrt{7} \times \sqrt{3}$ CDW structure. (e) and (f) are Fermi surfaces of monolayer VSe$_2$ under ±6% in-plane biaxial strain. First BZs are indicated by pink solid lines. The nesting vectors under good (poor) nesting conditions are denoted by purple (orange) arrows.

The effects of in-plane biaxial strain on the CDW instabilities in monolayer VSe$_2$ are also investigated. As shown in Fig. 4(a), compressive strains further soften the imaginary phonon modes at $\mathbf{q_1}$ and $\mathbf{q_3}$, indicating that both the $4 \times 4$ and $\sqrt{7} \times \sqrt{3}$ CDW instabilities have been enhanced. An additional Kohn anomaly appears near $\mathbf{q_6} = \frac{3}{8}\Gamma K$ point [see the corresponding

CDW vector in Fig. s8(b)], corresponding to one of the $4 \times \sqrt{3}$ CDW nesting vectors [see the structure and simulated STM of $4 \times \sqrt{3}$ CDW in Figs. 4(c) and s2(d), respectively]. These results suggest a more complicated competition among three CDW phases in VSe$_2$ monolayer by applying compressive strains. On the other hand, tensile strains push the lowest acoustic phonon band downwards, forming wider and deeper digs in the BZ [see Fig. 4(b)]. For the case of 6% tensile strain, the structure becomes very unstable with the appearance of the extended imaginary acoustic branch.

To study the CDW phase transitions by strain, the formation energies are also calculated for the three possible CDW structures under different strains. As shown in Fig. 4(d), both the compressive and tensile strains overall stabilize all the CDW structures, in contrast with the effects of charge doping. The $4 \times \sqrt{3}$ CDW structure has higher total energy than the other two CDW structures in the range of strain studied in this work. Under tensile strain and without strain, the $\sqrt{7} \times \sqrt{3}$ CDW phase has the lowest energy, while $4 \times 4$ CDW phase becomes more stable when compressive strain at and larger than 2%. These results indicate a phase transition between these two CDW states under compressive strain. The $4 \times 4$ CDW phase was observed experimentally based on the analysis of reflection high-energy electron diffraction (RHEED) and was proposed to have a small in-plane lattice constant (~3.31Å) due to the effect of substrate. [38] This is in line with our prediction. In addition, Zhang *et al.* had observed the $4 \times \sqrt{3}$ CDW structure in VSe$_2$ films [45] (80 layers thickness), and they attributed this new kind of CDW structure to the tensile strain (~4.5%) induced by the sapphire substrate. This controversial result may be caused by the difference between the phonon spectra of bulk and monolayer VSe$_2$ [46], attributed to the dimensionality change from 3D to 2D of the Fermi surface [18].

Figures 4(e) and 4(f) show the Fermi surface of normal VSe$_2$ under in-plane biaxial ±6% strain. The 6% compressive strain induces two nearly merged hole pockets at $\Gamma$ point, and triangular shape hole pockets centered at $K$ points [Fig. 4(e)]. It is interesting to note that such a Fermi surface topology can supply the nesting condition for three different CDWs: $4 \times 4$ ($\mathbf{q_1}$), $\sqrt{7} \times \sqrt{3}$ ($\mathbf{q_2}$ and $\mathbf{q_3}$), and $4 \times \sqrt{3}$ ($\mathbf{q_6}$) [Fig. s7(a)], consistent with the analysis of the phonon dispersion and formation energies discussed above [see Figs. 4(a) and 4(d)]. The 6% tensile strain widens cigar-like electron pockets at $M$ points, which makes the nesting condition for the $\sqrt{7} \times \sqrt{3}$ CDW unsatisfied [see the orange arrows shown in Fig. 4(f)]. Considering the broadened distribution of the imaginary frequencies along the $\Gamma$-$M$-$K$ path, we speculate that monolayer VSe$_2$ is dynamically unstable under such a large tensile strain.

**Conclusion**

In summary, first-principles calculations were performed to study the CDW phase transitions of monolayer VSe$_2$ by charge doping and in-plane biaxial strain. In the pristine VSe$_2$, the $\sqrt{7} \times \sqrt{3}$ CDW state is the ground state, in line with the experimental observations. However, it undergoes a phase transition to the $4 \times 4$ CDW state by applying larger than 2% compressive strain or doping with more than 0.25 electrons/f.u. and to the $3 \times \sqrt{3}$ CDW state by hole doping at a concentration at and above 0.2 holes/f.u.. Although electron-phonon coupling may play a crucial role, nesting effects were shown to be sufficient to explain most of the CDW phase transitions studied in this work except the case with 0.3 holes/f.u. doping. This indicates Fermi surface nesting is a good and simple starting point for understanding CDWs in monolayer VSe$_2$ [40]. This comprehensive study on the possible CDW states in monolayer VSe$_2$ suggests tunable transitions of the CDW states in monolayer VSe$_2$ by engineering which is appealing for applications in future electronic devices based on controllable multiple CDW phase transitions.

**Acknowledgements**

The authors would like to thank Dr. Chuan Chen and Dr. Yang-Hao Chan for the helpful discussions. The computational works were supported by the National Supercomputing Centre (NSCC) Singapore and Centre of Advanced 2D Materials (CA2DM) HPC infrastructure. This research is supported by the Ministry of Education, Singapore, under its MOE AcRF Tier 2 Award MOE2019-T2-2-30 and MOE AcRF Tier 1 Awards R-144-000-441-114 & R-144-000-413-114.

**References**

[1]  I. V. G. and A. A. F. K. S. Novoselov, A. K. Geim, S. V. Morozov, D. Jiang, Y. Zhang, S. V. Dubonos, *Electric Field Effect in Atomically Thin Carbon Films*, Science (80-. ). **306**, 666 (2004).

[2]  S. Manzeli, D. Ovchinnikov, D. Pasquier, O. V Yazyev, and A. Kis, *2D Transition Metal Dichalcogenides*, Nat. Rev. Mater. **2**, (2017).

[3]  D. Geng and H. Y. Yang, *Recent Advances in Growth of Novel 2D Materials: Beyond Graphene and Transition Metal Dichalcogenides*, Adv. Mater. **30**, 1 (2018).

[4]  M. Gibertini, M. Koperski, A. F. Morpurgo, and K. S. Novoselov, *Magnetic 2D Materials and Heterostructures*, Nat. Nanotechnol. **14**, 408 (2019).

[5] Kazuko Motizuki, *Structural Phase Transitions in Layered Transition Metal Compounds*, Vol. 8 (1986).

[6] A. F. Kusmartseva, B. Sipos, H. Berger, L. Forró, and E. Tutiš, *Pressure Induced Superconductivity in Pristine 1T-TiSe$_2$*, Phys. Rev. Lett. **103**, 236401 (2009).

[7] A. Rycerz, J. Tworzydło, and C. W. J. Beenakker, *Valley Filter and Valley Valve in Graphene*, Nat. Phys. **3**, 172 (2007).

[8] L. Xu, M. Yang, L. Shen, J. Zhou, T. Zhu, and Y. P. Feng, *Large Valley Splitting in Monolayer WS$_2$ by Proximity Coupling to an Insulating Antiferromagnetic Substrate*, Phys. Rev. B **97**, 041405(R) (2018).

[9] L. J. Li, E. C. T. O'Farrell, K. P. Loh, G. Eda, B. Özyilmaz, and A. H. Castro Neto, *Controlling Many-Body States by the Electric-Field Effect in a Two-Dimensional Material*, Nature **529**, 185 (2016).

[10] G. Gruner, *The Dynamics of Charge-Density Waves*, Rev. Mod. Phys. **60**, (1988).

[11] X. Zhu, Y. Cao, J. Zhang, E. W. Plummer, and J. Guo, *Classification of Charge Density Waves Based on Their Nature*, Proc. Natl. Acad. Sci. U. S. A. **112**, 2367 (2015).

[12] M. D. Johannes and I. I. Mazin, *Fermi Surface Nesting and the Origin of Charge Density Waves in Metals*, Phys. Rev. B - Condens. Matter Mater. Phys. **77**, (2008).

[13] J. Diego, A. H. Said, S. K. Mahatha, R. Bianco, L. Monacelli, M. Calandra, F. Mauri, K. Rossnagel, I. Errea, and S. Blanco-Canosa, *Van Der Waals Driven Anharmonic Melting of the 3D Charge Density Wave in VSe$_2$*, Nat. Commun. 598 (2021).

[14]  M. H. Whangbo and E. Canadell, *Analogies between the Concepts of Molecular Chemistry and Solid-State Physics Concerning Structural Instabilities. Electronic Origin of the Structural Modulations in Layered Transition-Metal Dichalcogenides*, J. Am. Chem. Soc. **114**, 9587 (1992).

[15]  A. W. Tsen, R. Hovden, D. Wang, Y. D. Kim, K. A. Spoth, Y. Liu, W. Lu, Y. Sun, J. C. Hone, L. F. Kourkoutis, P. Kim, and A. N. Pasupathy, *Structure and Control of Charge Density Waves in Two-Dimensional 1T-TaS$_2$*, Proc. Natl. Acad. Sci. U. S. A. **112**, 15054 (2015).

[16]  A. Kogar, G. A. De La Pena, S. Lee, Y. Fang, S. X. L. Sun, D. B. Lioi, G. Karapetrov, K. D. Finkelstein, J. P. C. Ruff, P. Abbamonte, and S. Rosenkranz, *Observation of a Charge Density Wave Incommensuration Near the Superconducting Dome in Cu$_x$TiSe$_2$*, Phys. Rev. Lett. **118**, (2017).

[17]  S. Gao, F. Flicker, R. Sankar, H. Zhao, Z. Ren, B. Rachmilowitz, S. Balachandar, F. Chou, K. S. Burch, Z. Wang, J. van Wezel, and I. Zeljkovic, *Atomic-Scale Strain Manipulation of a Charge Density Wave*, Proc. Natl. Acad. Sci. U. S. A. **115**, 6986 (2018).

[18]  P. Chen, W. W. Pai, Y. H. Chan, V. Madhavan, M. Y. Chou, S. K. Mo, A. V. Fedorov, and T. C. Chiang, *Unique Gap Structure and Symmetry of the Charge Density Wave in Single-Layer VSe$_2$*, Phys. Rev. Lett. **121**, 196402 (2018).

[19]  A. O. Fumega, M. Gobbi, P. Dreher, W. Wan, C. González-Orellana, M. Peña-Díaz, C. Rogero, J. Herrero-Martín, P. Gargiani, M. Ilyn, M. M. Ugeda, V. Pardo, and S. Blanco-

Canosa, *Absence of Ferromagnetism in VSe$_2$ Caused by Its Charge Density Wave Phase*, J. Phys. Chem. C **123**, (2019).

[20] M. Bonilla, S. Kolekar, Y. Ma, H. C. Diaz, V. Kalappattil, R. Das, T. Eggers, H. R. Gutierrez, M. H. Phan, and M. Batzill, *Strong Roomerature Ferromagnetism in VSe$_2$ Monolayers on van Der Waals Substrates*, Nat. Nanotechnol. **13**, 289 (2018).

[21] P. Coelho, K. Nguyen-Cong, M. Bonilla, S. K. Kolekar, M.-H. Phan, J. Avila, M. C. Asensio, I. I. Oleynik, and M. Batzill, *Charge Density Wave State Suppresses Ferromagnetic Ordering in VSe$_2$ Monolayers*, J. Phys. Chem. C **123**, 14089 (2019).

[22] P. K. J. Wong, W. Zhang, F. Bussolotti, X. Yin, T. S. Herng, L. Zhang, Y. L. Huang, G. Vinai, S. Krishnamurthi, D. W. Bukhvalov, Y. J. Zheng, R. Chua, A. T. N'Diaye, S. A. Morton, C. Y. Yang, K. H. Ou Yang, P. Torelli, W. Chen, K. E. J. Goh, J. Ding, M. T. Lin, G. Brocks, M. P. de Jong, A. H. Castro Neto, and A. T. S. Wee, *Evidence of Spin Frustration in a Vanadium Diselenide Monolayer Magnet*, Adv. Mater. **31**, (2019).

[23] V. N. Strocov, M. Shi, M. Kobayashi, C. Monney, X. Wang, J. Krempasky, T. Schmitt, L. Patthey, H. Berger, and P. Blaha, *Three-Dimensional Electron Realm in VSe2 by Soft-x-Ray Photoelectron Spectroscopy: Origin of Charge-Density Waves*, Phys. Rev. Lett. **109**, 086401 (2012).

[24] G. Duvjir, B. K. Choi, I. Jang, S. Ulstrup, S. Kang, T. Thi Ly, S. Kim, Y. H. Choi, C. Jozwiak, A. Bostwick, E. Rotenberg, J. G. Park, R. Sankar, K. S. Kim, J. Kim, and Y. J. Chang, *Emergence of a Metal-Insulator Transition and High-Temperature Charge-Density Waves in VSe$_2$ at the Monolayer Limit*, Nano Lett. **18**, 5432 (2018).

[25] G. Kresse and J. Hafner, *Ab Initio Molecular-Dynamics Simulation of the Liquid-Metal–Amorphous-Semiconductor Transition in Germanium*, Phys. Rev. B **49**, 14251 (1996).

[26] G. Kresse and J. Furthmüller, *Efficient Iterative Schemes for Ab Initio Total-Energy Calculations Using a Plane-Wave Basis Set*, Phys. Rev. B - Condens. Matter Mater. Phys. **54**, 11169 (1996).

[27] P.E.Blöchl, *Projector Augmented-Wave Method*, Phys. Rev. B **50**, 17953 (1994).

[28] J. P. Perdew, K. Burke, and M. Ernzerhof, *Generalized Gradient Approximation Made Simple*, Phys. Rev. Lett. **77**, 3865 (1996).

[29] W. Yu, J. Li, T. S. Herng, Z. Wang, X. Zhao, X. Chi, W. Fu, I. Abdelwahab, J. Zhou, J. Dan, Z. Chen, Z. Chen, Z. Li, J. Lu, S. J. Pennycook, Y. P. Feng, J. Ding, and K. P. Loh, *Chemically Exfoliated VSe$_2$ Monolayers with Room-Temperature Ferromagnetism*, Adv. Mater. **31**, 1903779 (2019).

[30] P. Giannozzi, S. Baroni, N. Bonini, M. Calandra, R. Car, C. Cavazzoni, D. Ceresoli, G. L. Chiarotti, M. Cococcioni, I. Dabo, A. Dal Corso, S. De Gironcoli, S. Fabris, G. Fratesi, R. Gebauer, U. Gerstmann, C. Gougoussis, A. Kokalj, M. Lazzeri, L. Martin-Samos, N. Marzari, F. Mauri, R. Mazzarello, S. Paolini, A. Pasquarello, L. Paulatto, C. Sbraccia, S. Scandolo, G. Sclauzero, A. P. Seitsonen, A. Smogunov, P. Umari, and R. M. Wentzcovitch, *QUANTUM ESPRESSO: A Modular and Open-Source Software Project for Quantum Simulations of Materials*, J. Phys. Condens. Matter **21**, 395502 (2009).

[31] P. Giannozzi, O. Andreussi, T. Brumme, O. Bunau, M. Buongiorno Nardelli, M. Calandra, R. Car, C. Cavazzoni, D. Ceresoli, M. Cococcioni, N. Colonna, I. Carnimeo, A. Dal

Corso, S. De Gironcoli, P. Delugas, R. A. Distasio, A. Ferretti, A. Floris, G. Fratesi, G. Fugallo, R. Gebauer, U. Gerstmann, F. Giustino, T. Gorni, J. Jia, M. Kawamura, H. Y. Ko, A. Kokalj, E. Küçükbenli, M. Lazzeri, M. Marsili, N. Marzari, F. Mauri, N. L. Nguyen, H. V. Nguyen, A. Otero-De-La-Roza, L. Paulatto, S. Poncé, D. Rocca, R. Sabatini, B. Santra, M. Schlipf, A. P. Seitsonen, A. Smogunov, I. Timrov, T. Thonhauser, P. Umari, N. Vast, X. Wu, and S. Baroni, *Advanced Capabilities for Materials Modelling with Quantum ESPRESSO*, J. Phys. Condens. Matter **29**, (2017).

[32] M. Esters, R. G. Hennig, and D. C. Johnson, *Dynamic Instabilities in Strongly Correlated VSe$_2$ Monolayers and Bilayers*, Phys. Rev. B **96**, 235147 (2017).

[33] P. V. C. Medeiros, S. Stafström, and J. Björk, *Effects of Extrinsic and Intrinsic Perturbations on the Electronic Structure of Graphene: Retaining an Effective Primitive Cell Band Structure by Band Unfolding*, Phys. Rev. B **89**, 041407(R) (2014).

[34] P. V. C. Medeiros, S. S. Tsirkin, S. Stafström, and J. Björk, *Unfolding Spinor Wave Functions and Expectation Values of General Operators: Introducing the Unfolding-Density Operator*, Phys. Rev. B **91**, 041116(R) (2015).

[35] D. L. Duong, M. Burghard, and J. C. Schön, *Ab Initio Computation of the Transition Temperature of the Charge Density Wave Transition in TiSe$_2$*, Phys. Rev. B - Condens. Matter Mater. Phys. **92**, 245131 (2015).

[36] W. Kohn, *Image of the Fermi Surface in the Vibration Spectrum of a Metal*, Phys. Rev. Lett. **2**, 393 (1959).

[37] W. Jolie, T. Knispel, N. Ehlen, K. Nikonov, C. Busse, A. Grüneis, and T. Michely, *Charge Density Wave Phase of VSe$_2$ Revisited*, Phys. Rev. B **99**, 115417 (2019).

[38] J. Feng, D. Biswas, A. Rajan, M. D. Watson, F. Mazzola, O. J. Clark, K. Underwood, I. Marković, M. McLaren, A. Hunter, D. M. Burn, L. B. Duffy, S. Barua, G. Balakrishnan, F. Bertran, P. Le Fèvre, T. K. Kim, G. Van Der Laan, T. Hesjedal, P. Wahl, and P. D. C. King, *Electronic Structure and Enhanced Charge-Density Wave Order of Monolayer VSe$_2$*, Nano Lett. **18**, 4493 (2018).

[39] J. G. Si, W. J. Lu, H. Y. Wu, H. Y. Lv, X. Liang, Q. J. Li, and Y. P. Sun, *Origin of the Multiple Charge Density Wave Order in 1T-VSe$_2$*, Phys. Rev. B **101**, 235405 (2020).

[40] D. Pasquier and O. V. Yazyev, *Unified Picture of Lattice Instabilities in Metallic Transition Metal Dichalcogenides*, Phys. Rev. B **100**, 201103(R) (2019).

[41] W. L. McMillan, *Theory of Discommensurations and the Commensurate-Incommensurate Charge-Density-Wave Phase Transition*, Phys. Rev. B **14**, 1496 (1976).

[42] D. Pasquier and O. V Yazyev, *Charge Density Wave Phase, Mottness, and Ferromagnetism in Monolayer 1T-NbSe$_2$*, Phys. Rev. B **98**, 045114 (2018).

[43] F. Weber, S. Rosenkranz, J. P. Castellan, R. Osborn, R. Hott, R. Heid, K. P. Bohnen, T. Egami, A. H. Said, and D. Reznik, *Extended Phonon Collapse and the Origin of the Charge-Density Wave in 2H-NbSe$_2$*, Phys. Rev. Lett. **107**, (2011).

[44] D. F. Shao, R. C. Xiao, W. J. Lu, H. Y. Lv, J. Y. Li, X. B. Zhu, and Y. P. Sun, *Manipulating Charge Density Waves in $1T-TaS_2$ by Charge-Carrier Doping: A First-Principles Investigation*, Phys. Rev. B **94**, 125126 (2016).

[45] D. Zhang, J. Ha, H. Baek, Y. H. Chan, F. D. Natterer, A. F. Myers, J. D. Schumacher, W. G. Cullen, A. V Davydov, Y. Kuk, M. Y. Chou, N. B. Zhitenev, and J. A. Stroscio, *Strain Engineering a 4a×√3a Charge-Density-Wave Phase in Transition-Metal Dichalcogenide 1 T-VSe$_2$*, Phys. Rev. Mater. **1**, 24005 (2017).

[46] It is noted that the position of the imaginary phonon corresponding to the $4 \times \sqrt{3}$ CDW is along HA path in bulk VSe$_2$, which doesn't exist in the monolayer 2D Fermi surface. And our calculations show that the energy of $4 \times \sqrt{3}$ CDW structure is close to $4 \times 4$ CDW structure under the 4.5% tensile strain, agreeing with their energy calculations.